\documentclass[aps,pre,reprint, amsmath, amssymb, superscriptaddress]{revtex4-2}

\usepackage{bm}
\usepackage[retainorgcmds]{IEEEtrantools}
\usepackage{graphicx}
\usepackage{mathrsfs}
\usepackage{amsmath}
\usepackage{amssymb}
\usepackage{color}
\usepackage{amsfonts}
\usepackage{amsthm}
\usepackage{times,txfonts}
\usepackage{enumerate}
\usepackage{dsfont}
\usepackage{nicefrac}

\newcommand{\ud}{\,\mathrm{d}}

\newcommand{\tr}[1]{\mathrm{Tr}\left( #1 \right)}
\newcommand{\cov}[2]{\mathrm{Cov}\left( #1 , #2 \right)}
\newcommand{\var}[1]{\mathrm{Var}\left( #1 \right)}
\newcommand{\avg}[1]{\left\langle #1 \right\rangle}

\newlength{\eqboxstorage}

\begin{document}

\title{A family of thermodynamic uncertainty relations valid for general fluctuation theorems}
\date{\today}
\author{Andr\'e M. Timpanaro}
\email{a.timpanaro@ufabc.edu.br}
\affiliation{Universidade Federal do ABC,  09210-580 Santo Andr\'e, Brazil}
\begin{abstract}
Thermodynamic Uncertainty Relations (TURs) are relations that establish lower bounds for the relative fluctuations of thermodynamic quantities in terms of the statistics of the associated entropy production. In this work we derive a family of TURs that explores higher order moments of the entropy production and is valid in any situation a Fluctuation Theorem holds. The resulting bound holds in both classical and quantum regimes and can always be saturated. These TURs are shown in action for a two level system weakly coupled to a bath undergoing a non time-symmetric drive, where we can use the Tasaki-Crooks fluctuation theorem. Finally, we draw a connection between our TURs and the existence of correlations between the entropy production and the thermodynamic quantity under consideration.
\end{abstract}
\maketitle{}

%
%
%
%

\section{Introduction}
In the recent decades, miniaturization has led to the development of devices and thermal machines operating in a nanoscopic scale \cite{benenti2017fundamental, Hanggi2012RMP}. These include quantum dots \cite{josefsson2018quantum}, nano-junction thermoelectrics \cite{dubi2011colloquium} and molecular motors \cite{mol-motors-nature, mol-motors-che-soc} to name a few. When considering the thermodynamics of these devices there are three important aspects of such systems. The first one is that contrary to macroscopic thermodynamic systems the fluctuations play an important role, so quantities like work and exchanged heat must be treated stochastically. Secondly, these systems are inherently out-of-equilibrium and so their operation entails some entropy production. Finally, these systems operate at a scale where quantum effects might have an important role and so a full quantum treatment is necessary for their study.

The effect that fluctuations have in a device is to make it more unreliable. For example, if during a particular run of an experiment the heat dissipation becomes significantly larger than average, then the device might be damaged or have its operation disrupted. On the other hand, a larger entropy production in a thermal machine leads to a reduced efficiency \cite{Pietzonka}. A natural question is whether it is possible to reconcile small fluctuations with small entropy production by designing the device in a clever way. An important step in answering this question was the development of the Thermodynamic Uncertainty Relations (TUR) \cite{Barato2015,Pietzonka2015,Pietzonka2017,Pietzonka2017a,gingrich2016dissipation,Dechant2018,BaratoNJP2018,Holubec2018PRL,proesmans2017discrete}. These are relations that relate the entropy production of the system with the relative fluctuations of currents in that same system through an inequality. The original TUR \cite{Barato2015} was formulated for classical Markovian systems and reads

\begin{equation}
\frac{\var{J}}{\avg{J}^2} \geq \frac{2}{\avg{\Sigma}}
\label{eq:TUR-classico}
\end{equation}
where $J$ is a current integrated over time and $\Sigma$ is the corresponding entropy production. Note that because of the structure of the inequality, the entropy production imposes a lower bound to the relative fluctuations of $J$ and this bound diverges when the entropy production becomes smaller, meaning that there must always be a tradeoff between relative fluctuations and dissipation.

Fluctuation theorems (FT) \cite{Gallavotti1995b,Jarzynski1997a,Crooks1998,Piechocinska2000,Tasaki2000,Kurchan2000,Jarzynski2004a, Andrieux2009,Saito2008,Esposito2009,Campisi2011,Jarzynski2011a,Talkner-Hangi} are another important class of results about the fluctuations of an out-of-equilibrium system. They connect the statistics of two related processes, called the forwards and backwards processes. The backwards process can be roughly seen as the time reversal of the forwards process, in the sense that any external drives are reversed and any sequence of interactions is done in a reversed order. However the initial state of the reversed process does not need to be the final state of the forwards process and often thermal states are used as initial states for both processes.

Mathematically, a FT may take a detailed form:
\begin{equation}
\frac{P_F(\Sigma, \phi)}{P_B(-\Sigma, -\phi)} = e^{\Sigma} 
\label{eq:FT-basico}
\end{equation}
where $\Sigma$ is the entropy production in an experimental run of the forwards process, $\phi$ is an exchanged quantity (or potentially a vector of exchanged quantities) and $P_{F(B)}$ are the distributions for the forwards (backwards) process. Equation (\ref{eq:FT-basico}) also leads to the integral form

\begin{equation}
\avg{e^{-\Sigma}}_F = 1
\label{eq:FT-Jarz}
\end{equation}

Because equation (\ref{eq:TUR-classico}) is only valid for classical systems and can in fact be violated in quantum scenarios \cite{Ptaszynski2018a}, there has been an ongoing effort to find analogous relations valid for quantum systems. Among these efforts we can cite TURs similar to (\ref{eq:TUR-classico}) but with a narrower validity regime \cite{MacIeszczak2018,Guarnieri2019}, works generalizing the classical notion of dynamical activity \cite{Hasegawa-activity} and works connecting TURs and FTs \cite{Merhav2010, Hasegawa2019, Potts2019, Timpanaro2019, TUT, Francica}, which will be our focus in this work. Most of the efforts to find TURs starting from FTs are restricted to the case where $P_F = P_B$ (which includes the Exchange Fluctuation Theorems \cite{Jarzynski2004a,Andrieux2009,Saito2008, Esposito2009,Campisi2011}), with the notable exceptions of \cite{Potts2019} where Potts and Samuelson derive the TUR-like expression

\begin{equation}
\frac{\var{\phi}_F + \var{\phi}_B}{(\avg{\phi}_F + \avg{\phi}_B)^2} \geq \frac{1}{\exp\left(\frac{\avg{\Sigma}_F + \avg{\Sigma}_B}{2}\right) - 1}
\label{eq:TUR-Potts}
\end{equation}
and \cite{Francica} where Francica derives the bound

\begin{equation}
2\left(\frac{\avg{\phi^2}_F + \avg{\phi^2}_B}{(\avg{\phi}_F + \avg{\phi}_B)^2}\right) \geq f\left(\frac{\avg{\Sigma}_F + \avg{\Sigma}_B}{2}\right) + 1
\label{eq:TUR-Francica-base}
\end{equation}
where $f(x) = \text{csch}^2(g(x/2))$ and $g(x)$ is the function inverse of $x \tanh(x)$. This can be reworded as the TUR

\begin{equation}
\frac{\var{\phi}_F + \var{\phi}_B}{(\avg{\phi}_F + \avg{\phi}_B)^2} \geq \frac{1}{2}\left(f\left(\frac{\avg{\Sigma}_F + \avg{\Sigma}_B}{2}\right)- \left(\frac{\avg{\phi}_F - \avg{\phi}_B}{\avg{\phi}_F + \avg{\phi}_B}\right)^2\right)
\label{eq:TUR-Francica}
\end{equation}
which provides a tighter bound than (\ref{eq:TUR-Potts}) when $P_F \simeq P_B$ but can be negative when the two processes have very different averages. In the symmetrical case these two bounds recover the bounds in \cite{Hasegawa2019} and \cite{Timpanaro2019} respectively.

In this work, we advance these efforts by deriving the following family of TUR-like expressions from a general fluctuation theorem obeying (\ref{eq:FT-basico}), which is the main result of this paper:
\begin{IEEEeqnarray}{rCl}
\nonumber \epsilon_{\alpha}(\phi) & \equiv & \frac{(1-\alpha) \var{\phi}_F + \alpha \var{\phi}_B} {(\avg{\phi}_F + \avg{\phi}_B)^2} \geq \\[0.2cm]
& \geq & \frac{\alpha}{2\avg{(\coth(\Sigma/2)+1 - 2\alpha)^{-1}}_F} -\alpha(1-\alpha)
\label{eq:TUR-alpha}
\end{IEEEeqnarray}
where $0 \leq \alpha \leq 1$ is a parameter that can be changed to obtain different bounds. In particular, the case $\alpha = \nicefrac{1}{2}$ that allows for a comparison with (\ref{eq:TUR-Potts}) and (\ref{eq:TUR-Francica}) leads to

\begin{equation}
\frac{\var{\phi}_F + \var{\phi}_B}{(\avg{\phi}_F + \avg{\phi}_B)^2} \geq \frac{1}{2}\left(\frac{1}{\avg{\tanh\left(\nicefrac{\Sigma}{2}\right)}}_F - 1\right)
\label{eq:TUR-1/2}
\end{equation}
which also leads to the bound found in \cite{TUT} if we further impose that $P_F = P_B$. It is important to note that $\alpha$ does not have any physical counterpart. Instead, it is a parameter that changes how much weight we are giving the forwards and backwards processes. So for example, $\alpha = \nicefrac{1}{3}$ leads to the bound

\begin{equation}
\frac{2\var{\phi}_F + \var{\phi}_B}{(\avg{\phi}_F + \avg{\phi}_B)^2} \geq \frac{1}{2\avg{(\coth\left(\nicefrac{\Sigma}{2}\right) + \nicefrac{1}{3})^{-1}}}_F - \frac{2}{3}
\label{eq:TUR-1/3}
\end{equation}
while $\alpha = \nicefrac{2}{3}$ leads to

\begin{equation}
\frac{\var{\phi}_F + 2\var{\phi}_B}{(\avg{\phi}_F + \avg{\phi}_B)^2} \geq \frac{1}{\avg{(\coth\left(\nicefrac{\Sigma}{2}\right) - \nicefrac{1}{3})^{-1}}}_F - \frac{2}{3}
\label{eq:TUR-2/3}
\end{equation}
where we have given more weight to respectively the fluctuations in the forwards and backwards processes.

Finally, the bound in (\ref{eq:TUR-alpha}) becomes an indeterminacy when $\alpha \rightarrow 0$. However, taking the limit allows us to obtain the inequality

\begin{equation}
\epsilon_{0}(\phi) = \frac{\var{\phi}_F}{(\avg{\phi}_F + \avg{\phi}_B)^2} \geq \frac{1}{\avg{e^{-2\Sigma}}_F - 1}
\label{eq:TUR-0}
\end{equation}
that interestingly turns out to be equivalent to the positivity of the covariance matrix between $\phi$ and $e^{-\Sigma}$ in the forwards process, as we show later.

We find for every $\alpha \in [0,1]$ optimal distributions, that saturate these inequalities even if we impose as constraints the values of $\avg{\phi}_F$ and $\avg{\phi}_B$ as well as the marginal distribution $P_F(\Sigma)$. In a sense this means that these inequalities use all the information available in the marginal distribution $P_F(\Sigma)$, which can already be seen by the average in (\ref{eq:TUR-alpha}) taking into account higher order moments of the entropy production.

It should be noted that this is not the first work arriving at a TUR that can be saturated (other examples include \cite{Timpanaro2019, TUT}) nor the first work to consider higher moments (for example \cite{TUT}), but as far as we know no TUR with these properties has been found for the case where $P_F \neq P_B$. This includes important cases, like the Tasaki-Crooks Fluctuation Theorem \cite{Crooks1998,Tasaki2000,Talkner-Hangi} as well as fluctuation theorems including feedback control \cite{Sagawa-Ueda}. Furthermore, the TURs valid in the case where the forwards and backwards processes differ are restricted to bounds for $\epsilon_{\nicefrac{1}{2}}$, while the results in eqs (\ref{eq:TUR-alpha}) and (\ref{eq:TUR-0}) can bound arbitrary mixtures of forwards and backwards variances (as exemplified in (\ref{eq:TUR-1/3}) and (\ref{eq:TUR-2/3})).

The rest of the work is organized as follows: we first have an abridged derivation of our main result (\ref{eq:TUR-alpha}), valid for any fluctuation theorem obeying (\ref{eq:FT-basico}), followed by some direct comparisons with existing TURs. After that we consider a physical example using the Tasaki-Crooks theorem with some comparisons with other TURs included and finally we have a discussion, explaining why our TUR (as well as the TURs in eqs (\ref{eq:TUR-Potts}) and (\ref{eq:TUR-Francica})) must trivialize if $\phi$ and $\Sigma$ are independent. All detailed derivations omitted can be found in the appendixes.

%
%
%
%

\section{Derivation of our results}
The main idea behind the derivation is to find the distribution $P_F(\Sigma, \phi)$ obeying (\ref{eq:FT-basico}) that minimizes $(1-\alpha) \var{\phi}_F + \alpha \var{\phi}_B$, using as constraints the values of the averages $\avg{\phi}_F$ and $\avg{\phi}_B$ as well as the marginal distribution $P_F(\Sigma)$.

The first point is to note that since $\avg{\phi}_F$ and $\avg{\phi}_B$ are fixed then we can just minimize $(1-\alpha) \avg{\phi^2}_F + \alpha \avg{\phi^2}_B$. Secondly, the quantity to minimize and all constraints can be expressed in terms of the distribution for the forwards process. Namely, $\avg{\phi}_B = \avg{-\phi e^{-\Sigma}}_F$, $\avg{\phi^2}_B = \avg{\phi^2 e^{-\Sigma}}_F$ and the fact that there exists $P_B$ satisfying (\ref{eq:FT-basico}) is equivalent to $\avg{e^{-\Sigma}}_F = 1$. In other words, we want to find the distribution $P(\Sigma, \phi)$ minimizing $\avg{\phi^2\left(1 - \alpha + \alpha e^{-\Sigma}\right)}$, subject to the constraints $\avg{\phi} = \avg{\phi}_F$, $\avg{-\phi e^{-\Sigma}} = \avg{\phi}_B$ and $P(\Sigma) = P_F(\Sigma)$.

We start with a simpler problem, where we consider a fixed process obeying (\ref{eq:FT-basico}) with forwards distribution $P$ and we will find the measurable function $f(\Sigma, \phi)$ that minimizes $\avg{f^2\left(1 - \alpha + \alpha e^{-\Sigma}\right)}$, subject to the constraints $\avg{f} = \avg{\phi}_F$ and $\avg{-f e^{-\Sigma}} = \avg{\phi}_B$. Mathematically, we want to minimize the functional

\begin{equation}
\mathcal{F}[f] = \int P(\Sigma, \phi)f(\Sigma, \phi)^2\left(1 - \alpha + \alpha e^{-\Sigma}\right) \ud\phi\ud\Sigma
\label{eq:functional-var}
\end{equation}
subject to the constraints

\begin{equation}
\int P(\Sigma, \phi)f(\Sigma, \phi) \ud\phi\ud\Sigma = \avg{\phi}_F
\label{eq:functional-avF}
\end{equation}
\begin{equation}
\int P(\Sigma, \phi)f(\Sigma, \phi)e^{-\Sigma} \ud\phi\ud\Sigma = -\avg{\phi}_B
\label{eq:functional-avB}
\end{equation}
The functional in (\ref{eq:functional-var}) is convex for $0\leq \alpha \leq 1$, while the constraints in (\ref{eq:functional-avF}) and (\ref{eq:functional-avB}) are given by linear functionals, implying that $f$ can be easily found using Calculus of Variations. This leads to the following equation for the optimal $f$:

\begin{equation}
P(\Sigma, \phi)\left(f(\Sigma, \phi)\left(1 -\alpha + \alpha e^{-\Sigma}\right) - \left(\lambda + \mu e^{-\Sigma}\right)\right) = 0\,\,\forall\,\, \Sigma, \phi
\label{eq:f-eq}
\end{equation}
where $\lambda$ and $\mu$ are Lagrange multipliers that must be found using (\ref{eq:functional-avF}) and (\ref{eq:functional-avB}). This implies that
\begin{equation}
f(\Sigma, \phi) = \frac{\lambda + \mu e^{-\Sigma}}{1 -\alpha + \alpha e^{-\Sigma}}
\label{eq:f}
\end{equation}
for all points $(\Sigma, \phi)$ in the support of $P$. The important thing is that since $\phi$ itself is a measurable function of $(\Sigma, \phi)$ obeying the same constraints as $f$, then we have

\begin{IEEEeqnarray}{rCl}
\nonumber
\avg{f^2\left(1 - \alpha + \alpha e^{-\Sigma}\right)} & \leq &
\avg{\phi^2\left(1 - \alpha + \alpha e^{-\Sigma}\right)}_F = \\ [0.2cm]
&=& (1 - \alpha)\avg{\phi^2}_F + \alpha \avg{\phi^2}_B
\label{eq:inequality-moments}
\end{IEEEeqnarray}
which already encapsulates a TUR-like relation. In order to find it explicitly we have to find the values of the Lagrange multipliers, substitute in $f$ and then subtract $(1 - \alpha)\avg{\phi}_F^2 + \alpha \avg{\phi}_B^2$ in both sides of (\ref{eq:inequality-moments}). The detailed calculation is in appendix \ref{ap:deduction}, but the final results are that if $0 < \alpha \leq 1$, then

\begin{equation}
\epsilon_{\alpha}(\phi) \geq \frac{\alpha}{2\avg{(\coth(\Sigma/2)+1 - 2\alpha)^{-1}}_F} -\alpha(1-\alpha)
\label{eq:TUR-alpha-encore}
\end{equation}
and for $\alpha = 0$ we have

\begin{equation}
\epsilon_{0}(\phi) = \frac{\var{\phi}_F}{(\avg{\phi}_F + \avg{\phi}_B)^2} \geq \frac{1}{\avg{e^{-2\Sigma}}_F - 1}
\label{eq:TUR-0-encore}
\end{equation}

What we just did actually solves the problem of finding an optimal forwards distribution, since if $\lambda, \mu$ are the Lagrange multipliers corresponding to the optimal $f$, for given $\avg{\phi}_F, \avg{\phi}_B$ then a forwards distribution whose support obeys

\begin{equation}
\phi = \frac{\lambda + \mu e^{-\Sigma}}{1 -\alpha + \alpha e^{-\Sigma}}
\label{eq:opt-support}
\end{equation}
satisfies the constraints and saturates the inequality for the corresponding $\alpha$ (since we will have $f = \phi$). This saturating joint distribution $P_F(\Sigma, \phi)$ can be built for any desired marginal distribution $P_F(\Sigma)$ satisfying $\avg{e^{-\Sigma}}_F = 1$, with the exception of $P_F(\Sigma) = \delta(\Sigma)$ and cases where either $\avg{e^{-2\Sigma}}_F = \infty$ or $\avg{e^{\Sigma}}_F = \infty$ (these edge cases can still be saturated by a different construction, see appendix \ref{ap:saturation}).

%
%
%
%


\section{Saturation and Comparisons}

\begin{figure}[htb!]
\centering
\includegraphics[width=0.48\textwidth]{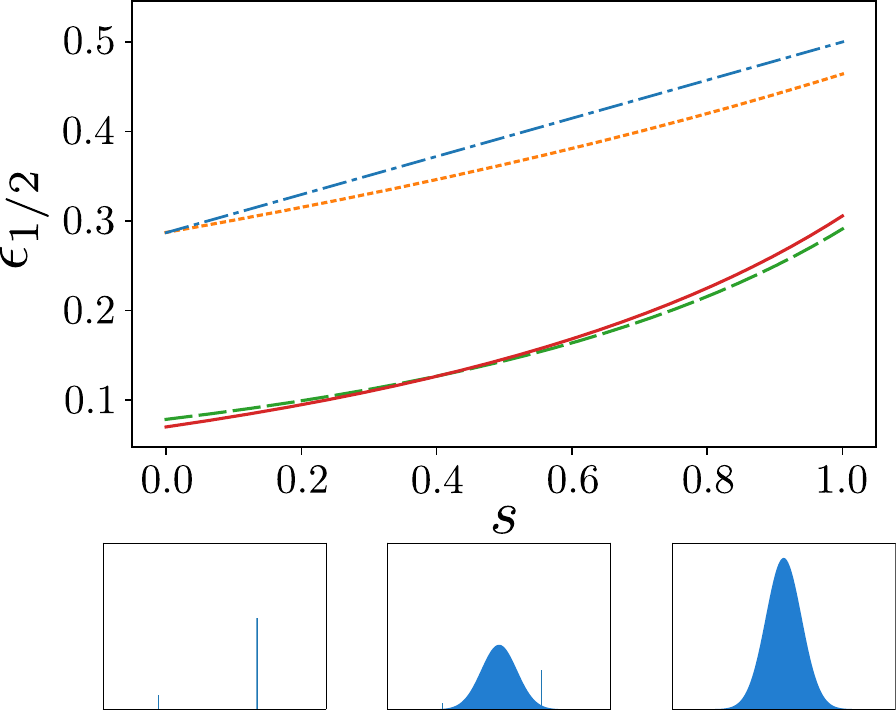}
\caption{\label{fig:compare} Comparison between the value of $\epsilon_{\nicefrac{1}{2}}(\phi)$ (in blue/dot-dashed), our bound (\ref{eq:TUR-1/2}) (in orange/dotted), the Potts-Samuelson bound (\ref{eq:TUR-Potts}) (in green/dashed) and the Francica bound (\ref{eq:TUR-Francica}) (in red/solid) for different distributions. The distribution $P_s$ (illustrated in the bottom panels) is given by $P_s = sP_1 + (1-s)P_0$, where $P_0$ is the (unique) distribution with two points in the support satisfying $\avg{\Sigma}_F=1, \avg{\Sigma}_B=3,\avg{\phi}_F=7$ and $\avg{\phi}_B=3$, and $P_1$ is the normal bivariate distribution satisfying $\avg{\Sigma}_F=1,\avg{\phi}_F=7, \var{\Sigma}_F=2,\cov{\Sigma}{\phi}_F=10, \var{\phi}_F=50.001$.
}
\end{figure}

To get an idea of how this new bound compares with the ones found in the literature, we will consider a family of distributions that interpolate between one having two points on its support and a bivariate normal distribution. Since the bounds in eqs (\ref{eq:TUR-Potts}) and (\ref{eq:TUR-Francica}) refer only to

\begin{equation}
    \frac{\var{\phi}_F + \var{\phi}_B}{(\avg{\phi}_F + \avg{\phi}_B)^2} = 2\epsilon_{\nicefrac{1}{2}}(\phi)
\end{equation}
then we need to restrict ourselves to the case $\alpha = \nicefrac{1}{2}$. The graph showing the actual value of $\epsilon_{\nicefrac{1}{2}}(\phi)$, the values of the bounds, as well as the details of the distributions involved can be found in fig \ref{fig:compare}.

Note in fig \ref{fig:compare} that our bound saturates in the case of the distribution with two points in the support. The reason behind this is detailed in appendix \ref{ap:two-point}, but intuitively it is because eq (\ref{eq:opt-support}) has two free variables ($\lambda$ and $\mu$) that can be adjusted to fit whatever the coordinates of the two points are, meaning that a two point distribution naturally obeys eq (\ref{eq:opt-support}) which in itself means the bound saturates. As detailed in the appendix, the only exception is when $P_F(\Sigma) = \delta(\Sigma)$.

%
%
%
%

\section{Physical Example}
For our physical example, we consider a two level system weakly coupled to a bath at temperature $T$. The system is subject to a drive such that it has a time dependant hamiltonian given by

\begin{equation}
H(t) = 
\begin{bmatrix}
0 & a\sin(\Omega \,t) \\
\overline{a} \sin(\Omega \,t) & \quad E(1 + \cos(\Omega \,t))
\end{bmatrix}
\label{eq:drive}
\end{equation}
and the system is initially in thermal equilibrium with the bath:

\begin{equation}
\rho_0 = \frac{e^{-\beta H(0)}}{\tr{e^{-\beta H(0)}}}
\end{equation}
Our motivation behind this example is simply to provide a minimal system, where there exists and external drive that is not time symmetrical, implying the forwards and backwards processes have different statistics.

In this situation, the Tasaki-Crooks Fluctuation Theorem holds. Namely, if the forwards process is the evolution of the system under the drive until a time instant $t_f$, starting from the state $\rho_0$, then the backwards process is the evolution of the system under the hamiltonian $H^B(t) = H(t_f - t)$, starting from the state

\begin{equation}
\rho^B_0 = \frac{e^{-\beta H^B(0)}}{\tr{e^{-\beta H^B(0)}}} = \frac{e^{-\beta H(t_f)}}{\tr{e^{-\beta H(t_f)}}}
\end{equation}
and the work distribution is such that

\begin{figure}[htb!]
\centering
\includegraphics[width=0.48\textwidth]{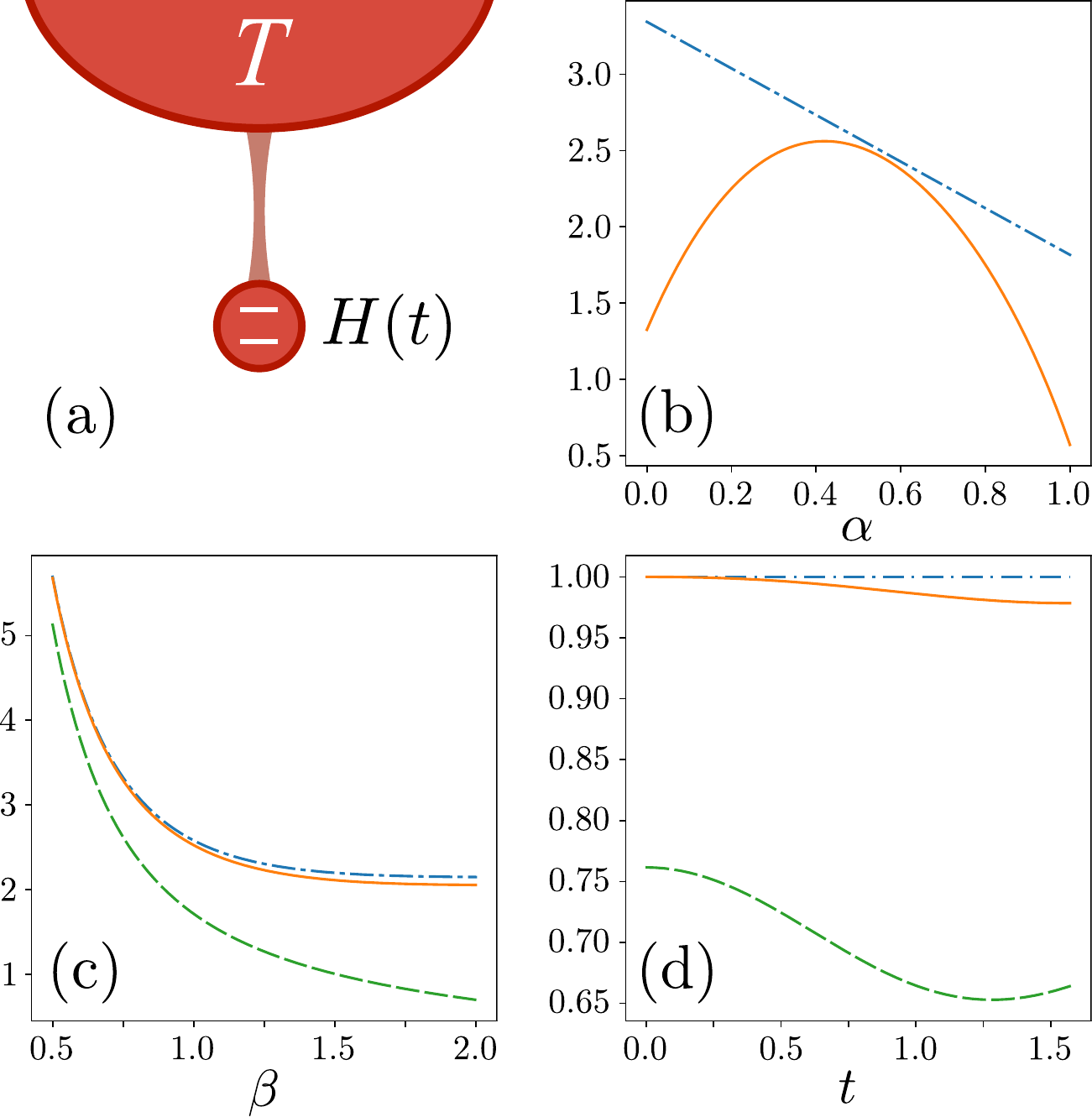}
\caption{\label{fig:example} Verification and comparison of our bound in a physical example. {\bf (a)} The setup considered: a qubit with a time dependant hamiltonian $H(t)$, as given by equation (\ref{eq:drive}) weakly coupled to a bath at temperature $T$. In all simulations the parameters $E = 1$, $a = 1 + i$ and $\Omega = 1$ were used. {\bf (b)} In this case $\beta = 1$ and $t_f = \nicefrac{\pi}{2}$ were used. The graph shows the relative fluctuations $\epsilon_{\alpha}(W)$ (blue/dot-dashed curve) compared with our lower bound in equation (\ref{eq:TUR-alpha}) (orange/solid curve) for different values of $\alpha$ {\bf (c)} In this case $t_f = \nicefrac{\pi}{2}$, $\beta$ is varied from $0.5$ to 2 and only $\alpha = \nicefrac{1}{2}$ is considered. The graph shows the relative fluctuations $\epsilon_{\nicefrac{1}{2}}(W)$ (blue/dot-dashed curve), our bound (orange/solid curve) and for comparison the maximum between the bounds in eqs (\ref{eq:TUR-Potts}) and (\ref{eq:TUR-Francica}) (green/dashed curve) {\bf (d)} Time evolution of the fluctuations and the bounds \--- Here we consider $\beta = 1$, $\alpha = \nicefrac{1}{2}$ and $t_f$ varied from 0 to $\nicefrac{\pi}{2}$. The order of magnitude of fluctuations changes wildly, so we look instead to the ratios of $\epsilon_{\nicefrac{1}{2}}(W)$ with our bound (orange/solid curve) and the maximum between the bounds in eqs (\ref{eq:TUR-Potts}) and (\ref{eq:TUR-Francica}) (green/dashed curve). The blue/dot-dashed line at 1 would mark the saturation of the bound.}
\end{figure}

\begin{equation}
\frac{P_F(W)}{P_B(-W)} = e^{\beta(W - \Delta F)}
\label{eq:crooks}
\end{equation}
where

\begin{equation}
\Delta F = \frac{1}{\beta}\log\left(\frac{\tr{e^{-\beta H(0)}}}{\tr{e^{-\beta H(t_f)}}}\right)\quad\mbox{and}\quad \Sigma = \beta(W - \Delta F)
\end{equation}
Since the entropy production in the forwards process is a direct function of the work in the same process and $\Delta F$ changes signs if we switch the forwards and backwards processes, then the equality (\ref{eq:crooks}) can indeed be put in the form of equation (\ref{eq:FT-basico}):

\begin{equation}
\frac{P_F(\Sigma, W)}{P_B(-\Sigma, -W)} = e^{\Sigma}
\label{eq:crooks-2}
\end{equation}
To find the work distributions for the evolution up to time $t_f$, we follow \cite{Talkner-Hangi} and use the characteristic function

\begin{equation}
\chi_F(z;t_f) = \tr{U^{\dagger}e^{izH(t_f)}Ue^{-izH(0)}\rho_0}
\end{equation}
\begin{equation}
\chi_B(z;t_f) = \tr{Ue^{izH(0)}U^{\dagger}e^{-izH(0)}\rho^B_0}
\end{equation}
where $U$ is the unitary operator giving the forwards evolution from the starting state to the final state. Since the drive in (\ref{eq:drive}) is not time symmetrical, then in general we have $U \neq U^{\dagger}$ and $P_F \neq P_B$. In order to compare with the TUR inequalities in \cite{Potts2019, Proesmans} we must consider the case $\alpha = \nicefrac{1}{2}$ in equation (\ref{eq:TUR-1/2}). 
Figure \ref{fig:example} contains graphs verifying our bound and comparing it to the ones in \cite{Potts2019, Proesmans} and \cite{Francica} as $\alpha$, $\beta$ and $t_f$ are changed. In these graphs it becomes clear the advantage of using higher moments in order to obtain a more precise bound and particularly in figure \ref{fig:example}c, our bound is very close to being saturated in the whole range tested.

%
%
%
%

\section{The role of correlations}
An interesting point about our bounds (\ref{eq:TUR-alpha}) and (\ref{eq:TUR-0}), as well as the earlier bound (\ref{eq:TUR-Potts}) is that they become trivial in the case $\avg{\phi}_F + \avg{\phi}_B = 0$. However we have

\begin{equation}
\cov{\phi}{e^{-\Sigma}}_F = \avg{\phi e^{-\Sigma}}_F - \avg{\phi}_F\avg{e^{-\Sigma}}_F
 = -\left(\avg{\phi}_F + \avg{\phi}_B\right)
\label{eq:cov}
\end{equation}

That is, the bound in equation (\ref{eq:TUR-alpha}) can be rewritten as

\begin{IEEEeqnarray}{C}
(1-\alpha) \var{\phi}_F + \alpha \var{\phi}_B \geq  \\[0.2cm]
\nonumber \geq \left(\frac{\alpha}{2\avg{(\coth(\Sigma/2)+1 - 2\alpha)^{-1}}_F} -\alpha(1-\alpha)\right)\cov{\phi}{e^{-\Sigma}}^2_F
\label{eq:TUR-alpha-cov} 
\end{IEEEeqnarray}
and hence the bounds trivialize when $\phi$ and $\Sigma$ are uncorrelated. This can be better understood, by checking the solution for $f(\Sigma, \phi)$ in equation (\ref{eq:f}). The full calculation in appendix \ref{ap:deduction} shows that we must have in this case

\begin{equation}
f(\Sigma, \phi) = \avg{\phi}_F
\end{equation}
from which one can easily check that $\avg{f}_{F} = \avg{\phi}_{F}$ and $\avg{-f e^{-\Sigma}}_{F} = -\avg{\phi}_{F} = \avg{\phi}_{B}$. In other words, a constant $\phi$ is consistent with our constraints in this case. Note that this is in contrast with the more studied case $P_F = P_B$ where $\phi$ can only be a constant if the averages are 0. This is a direct consequence of $\avg{\phi}$ being the same in the forwards and backwards process in this case, meaning by equation (\ref{eq:cov}) that there is always a correlation between $\phi$ and $\Sigma$ when the forwards and backwards processes coincide, leading to non-trivial bounds.

Another point relating specifically to the case $\alpha = 0$ is that

\begin{equation}
\var{e^{-\Sigma}}_F = \avg{e^{-2\Sigma}}_F - \avg{e^{\Sigma}}_F^2 = \avg{e^{-2\Sigma}}_F - 1
\end{equation}
meaning the inequality for $\alpha = 0$ can be rewritten as 

\begin{equation}
\var{\phi}_F \var{e^{-\Sigma}}_F \geq \cov{\phi}{e^{-\Sigma}}^2_F
\end{equation}
which follows directly from the positivity of the covariance matrix between $\phi$ and $e^{-\Sigma}$ for the forwards process (by symmetry, the case $\alpha = 1$ has the same interpretation, but for the backwards process). Note that even though this makes these cases look trivial, it is still non trivial that the inequality can be saturated.

%
%
%
%

\section{Conclusions}
In this work, we have presented a rigorous derivation of a family of thermodynamic uncertainty relations that are direct consequences of fluctuation theorems. Contrary to most similar works, we do not require the forwards and backwards processes to have the same statistics, allowing the application in situations where the Tasaki-Crooks FT holds and situations in the presence of feedback control. Also, contrary to previously known bounds the present result can always be saturated, even when the two processes differ.

The bound is obtained by an optimization procedure and the result can be interpreted as the tightest bound given as constraints the marginal distribution for entropy production and the forwards and backwards averages of the current whose fluctuations we want to bound. We are able to find explicitly which joint distributions (or sequences of joint distributions, as is necessary in some edge cases) saturate the bound, showing that this depends only on what their support is.

Finally, we show that our bounds, as well as other bounds known for the case where the forwards and backwards processes are distinct must trivialize when the current being bound is uncorrelated with the entropy production, giving some insight on the connection between TURs and correlations. More precisely, this leads to an interpretation where TURs derived from fluctuation theorems are deep down statements about the correlations between the entropy production and other currents.


\bibliography{library}

\pagebreak
\widetext

\appendix
\onecolumngrid

\setcounter{equation}{0}
\setcounter{figure}{0}
\setcounter{table}{0}

\section{Deduction of the TUR family}
\label{ap:deduction}

Like in the main text, we consider a distribution $P(\Sigma, \phi)$ for the forwards process and want to find the measurable function $f(\Sigma, \phi)$ that minimizes the functional

\begin{equation}
\mathcal{F}[f] = \int P(\Sigma, \phi)f(\Sigma, \phi)^2\left(1 - \alpha + \alpha e^{-\Sigma}\right) \ud\phi\ud\Sigma
\label{ap:eq:functional-var}
\end{equation}
subject to the constraints

\begin{equation}
\int P(\Sigma, \phi)f(\Sigma, \phi) \ud\phi\ud\Sigma = \avg{\phi}_F
\label{ap:eq:functional-avF}
\end{equation}
\begin{equation}
\int P(\Sigma, \phi)f(\Sigma, \phi)e^{-\Sigma} \ud\phi\ud\Sigma = -\avg{\phi}_B
\label{ap:eq:functional-avB}
\end{equation}
In order to do this minimization, we must consider the following Lagrangian functional

\begin{equation}
\mathcal{L}[f] = \int P(\Sigma, \phi)\left(f(\Sigma, \phi)^2\left(1 - \alpha + \alpha e^{-\Sigma}\right) + \lambda\left(f(\Sigma, \phi) - \avg{\phi}_F\right) + \mu\left(f(\Sigma, \phi)e^{-\Sigma} + \avg{\phi}_B\right) \right) \ud\phi\ud\Sigma
\end{equation}
Since $\mathcal{L}$ is convex whenever $0 \leq \alpha \leq 1$ and we have only linear equality constraints, then for these values of $\alpha$ we can find an equation for the minimum $f$ by simply equating the functional derivative to zero. Since

\begin{equation}
\frac{\delta \mathcal{L}}{\delta f} = P(\Sigma, \phi)\left(2f(\Sigma, \phi)\left(1 - \alpha + \alpha e^{-\Sigma}\right) + \lambda + \mu e^{-\Sigma}\right)
\end{equation}
then for all $(\Sigma, \phi)$ in the support we have

\begin{equation}
f(\Sigma, \phi) = \frac{-(\lambda + \mu e^{-\Sigma})}{2\left(1 - \alpha + \alpha e^{-\Sigma}\right)}
\label{ap:eq:f}
\end{equation}

The form of the solution is different for $\alpha = 0$ and $0 < \alpha \leq 1$. We will start with the simplest case:

\subsection{The $\alpha = 0$ case}

In this case eq (\ref{ap:eq:f}) can be simplified to

\begin{equation}
f(\Sigma, \phi) = A + B e^{-\Sigma}
\label{eq:f-alpha-0}
\end{equation}
Imposing the constraints leads to the system

\begin{equation}
\left\{
\begin{array}{l}
\avg{\phi}_F = \avg{A + B e^{-\Sigma}}_F = A + B \vspace{0.2cm} \\
\avg{\phi}_B = \avg{-e^{-\Sigma}\left(A + B e^{-\Sigma}\right)}_F = -A - B \avg{e^{-2\Sigma}}_F \\
\end{array}
\right.
\label{eq:sys-alpha-0}
\end{equation}
with solution

\begin{equation}
B = \frac{\avg{\phi}_F + \avg{\phi}_B}{1 - \avg{e^{-2\Sigma}}_F} \quad\quad\mbox{and}\quad\quad A = \avg{\phi}_F - B
\label{eq:sol-sys-alpha-0}
\end{equation}
implying that $\avg{f^2}_F$

\begin{IEEEeqnarray}{rCl}
\nonumber
\avg{f^2}_F &=& \int f(\Sigma, \phi)^2 P_F(\Sigma, \phi) \ud\Sigma \ud\phi \\[0.2cm]
\nonumber
&=& \int (A + B e^{-\Sigma})^2 P_F(\Sigma, \phi) \ud\Sigma \ud\phi \\[0.2cm]
\nonumber
&=& A^2 + 2AB + B^2 \avg{e^{-2\Sigma}}_F \\[0.2cm]
&=& (A + B)^2 + B^2 \left(\avg{e^{-2\Sigma}}_F - 1\right) \\[0.2cm]
&=& \avg{\phi}_F^2 + \frac{(\avg{\phi}_F + \avg{\phi}_B)^2}{\avg{e^{-2\Sigma}}_F - 1} \Rightarrow
\label{eq:phi2-alpha-0}
\end{IEEEeqnarray}

\begin{equation}
\var{f}_F = \avg{f^2}_F - \avg{f}_F^2 = \frac{(\avg{\phi}_F + \avg{\phi}_B)^2}{\avg{e^{-2\Sigma}}_F - 1}
\end{equation}
Since $\phi$ is a measurable function of $(\Sigma, \phi)$ satisfying the same constraints as $f$, it follows that $\var{\phi}_F \geq \var{f}_F$ and hence

\begin{equation}
\var{\phi}_F \geq \frac{(\avg{\phi}_F + \avg{\phi}_B)^2}{\avg{e^{-2\Sigma}}_F - 1}
\end{equation}

\subsection{The $0 < \alpha \leq 1$ case}

If $\alpha \neq 0$, then eq (\ref{ap:eq:f}) can always be rewritten as

\begin{equation}
f(\Sigma, \phi) = A + \frac{B}{\Omega(\Sigma)} \quad\quad\mbox{where}\quad\quad \Omega(\Sigma) = 1 - \alpha + \alpha e^{-\Sigma}
\end{equation}
Furthermore, if we define
\begin{equation}
\avg{\frac{1}{\Omega}}_F = \omega \quad\quad\mbox{and}\quad\quad \avg{\frac{e^{-\Sigma}}{\Omega}}_F = \omega'
\end{equation}
then we have

\begin{equation}
1 = \avg{\frac{1 - \alpha + \alpha e^{-\Sigma}}{\Omega}}_F = (1 - \alpha)\omega + \alpha \omega' \quad\Rightarrow\quad \omega' = \frac{1 - \omega}{\alpha} + \omega
\end{equation}

So imposing the constraints leads to the system

\begin{equation}
\left\{
\begin{array}{l}
\avg{\phi}_F = \avg{A + \nicefrac{B}{\Omega}}_F = A + B \omega \vspace{0.2cm} \\
\avg{\phi}_B = \avg{-e^{-\Sigma}\left(A + \nicefrac{B}{\Omega}\right)}_F = -A - B \omega' \\
\end{array}
\right.
\label{eq:sys-alpha-gen}
\end{equation}

with solution

\begin{equation}
B = \frac{\alpha(\avg{\phi}_F + \avg{\phi}_B)}{\omega - 1}\quad\quad\mbox{and}\quad\quad A = \avg{\phi}_F - B\omega
\end{equation}

Finally,
\begin{IEEEeqnarray}{rCl}
\nonumber
\avg{f^2\Omega}_F &=& \avg{\left(A + \frac{B}{\Omega}\right)^2 \Omega}_F \\[0.2cm]
\nonumber
&=& \avg{A^2\Omega + 2AB + \frac{B^2}{\Omega}}_F \\[0.2cm]
\nonumber
&=& A^2+ 2AB + B^2\omega \\[0.2cm]
\nonumber
&=& (A + B)^2 + B^2(\omega - 1) \\[0.2cm]
\nonumber
&=& (\avg{\phi}_F + B(1-\omega))^2 + B^2(\omega - 1) \\[0.2cm]
&=& \avg{\phi}_F^2 - 2\alpha\avg{\phi}_F(\avg{\phi}_F + \avg{\phi}_B) + \frac{\alpha^2(\avg{\phi}_F + \avg{\phi}_B)^2\omega}{\omega - 1}
\end{IEEEeqnarray}
leading to
\begin{IEEEeqnarray}{rCl}
\nonumber
(1-\alpha) \var{f}_F + \alpha \var{f}_B &=& \avg{f^2\Omega}_F - (1-\alpha)\avg{\phi}_F^2 - \alpha\avg{\phi}_B^2 \\[0.2cm]
\nonumber
 &=& \alpha\avg{\phi}_F^2 - 2\alpha\avg{\phi}_F(\avg{\phi}_F + \avg{\phi}_B) -\alpha\avg{\phi}_B^2 + \frac{\alpha^2(\avg{\phi}_F + \avg{\phi}_B)^2\omega}{\omega - 1} \\[0.2cm]
&=& (\avg{\phi}_F + \avg{\phi}_B)^2 \left(\frac{\alpha^2\omega}{\omega - 1} -\alpha\right)
\end{IEEEeqnarray}
and the inequality

\begin{equation}
\frac{(1-\alpha) \var{\phi}_F + \alpha \var{\phi}_B}{(\avg{\phi}_F + \avg{\phi}_B)^2} \geq \frac{\alpha^2\omega}{\omega - 1} -\alpha
\end{equation}
To get to the expression in the main text we still need to do some algebraic manipulations. We first note that

\begin{equation}
\frac{\alpha^2\omega}{\omega - 1} -\alpha = \frac{\alpha^2}{\omega - 1} - \alpha(1-\alpha)
\label{eq:simple-1}
\end{equation}
and that

\begin{equation}
\omega - 1 = \avg{\frac{1}{1-\alpha + \alpha e^{-\Sigma}}}_F - 1 = \alpha\,\avg{\frac{1-e^{-\Sigma}}{1-\alpha + \alpha e^{-\Sigma}}}_F
\label{eq:simple-2}
\end{equation}
Next, we note that

\begin{equation}
\frac{1-e^{-\Sigma}}{1-\alpha + \alpha e^{-\Sigma}} = \left(\frac{1-\alpha + \alpha e^{-\Sigma}}{1-e^{-\Sigma}}\right)^{-1} = \left(\frac{1}{1-e^{-\Sigma}}\,\, - \alpha\right)^{-1} = \left(\frac{\coth(\Sigma/2)+1}{2}\,\, - \alpha\right)^{-1} = \frac{2}{\coth(\Sigma/2)+1 - 2\alpha}
\label{eq:simple-3}
\end{equation}
Substituting (\ref{eq:simple-3}) in (\ref{eq:simple-2}) leads to

\begin{equation}
\omega - 1 = 2\alpha \avg{(\coth(\Sigma/2)+1 - 2\alpha)^{-1}}_F
\label{eq:simple-4}
\end{equation}
and finally substituting in (\ref{eq:simple-1}) we get

\begin{equation}
\frac{(1-\alpha) \var{\phi}_F + \alpha \var{\phi}_B}{(\avg{\phi}_F + \avg{\phi}_B)^2} \geq \frac{\alpha}{2\avg{(\coth(\Sigma/2)+1 - 2\alpha)^{-1}}_F} -\alpha(1-\alpha)
\end{equation}
From this point forward, the deduction in the main text is complete.

\subsection{Saturation}
\label{ap:saturation}
What we derived so far is already sufficient to show that the bound can be saturated {\bf when all averages are well behaved}. This is because the solution for $f$ is always only a function of $\Sigma$, so by considering the joint distribution of $\Sigma$ and $\phi' = f(\Sigma)$ we get a distribution that saturates the bound.

There are two possible technical problems that we ignored so far. The first one is that the system we use to determine $f$ (eqs (\ref{eq:sys-alpha-0}) and (\ref{eq:sys-alpha-gen})) may be linearly dependent. This happens for $\alpha = 0$ if $\avg{e^{-2\Sigma}}_F = 1$ and for $\alpha > 0$ if $\omega = 1$. In both of these cases we can use Jensen's inequality to get to the conclusion that $P_F(\Sigma) = \delta(\Sigma)$. The reasoning is as follows:

\[
\mbox{We have}\quad\quad
\avg{e^{-2\Sigma}}_F = \avg{\left(e^{-\Sigma}\right)^2}_F \geq \avg{e^{-\Sigma}}_F^2 = 1
\quad\quad\mbox{and}\quad\quad
\omega = \avg{\frac{1}{1-\alpha + \alpha e^{-\Sigma}}}_F \geq \frac{1}{\avg{1-\alpha + \alpha e^{-\Sigma}}_F} = 1.
\]
Since both $x^2$ and $\nicefrac{1}{x}$ are strictly convex functions of the positive reals, then saturation is only possible if the support is a singleton, which implies $\Sigma = 0$ since $\avg{e^{-\Sigma}}_F = 1$. Another consequence is that $\avg{\phi}_F + \avg{\phi}_B$ must be 0 because of the FT symmetry, so the bounds (as well as the $\epsilon_{\alpha}$) are undefined in this case. Nevertheless, this implies that the support for $\phi$ can also be picked as a singleton, so there exists a distribution with minimal variance satisfying the constraints.

The second problem is more complicated and arises only for $\alpha = 0$ and $\alpha = 1$. We will focus in the case $\alpha = 0$, since $\alpha = 1$ works the same way (switching the forwards and backwards processes). The difficulty is that we can have $\avg{e^{-2\Sigma}}_F = \infty$, trivializing the bound. However, in this case (\ref{eq:sol-sys-alpha-0}) would lead to $f(\Sigma, \phi) = \avg{\phi}_F$ which gives the wrong average for the backwards process in general.

What we can do in this case is build a sequence of distributions $P_{F,n}$ such that all $P_{F,n}$ satisfy the constraints and

\begin{equation}
\lim_{n\rightarrow \infty} \var{\phi}_{F,n} = 0
\end{equation}
so the bound can still be saturated asymptotically. The idea is to use eq (\ref{eq:f-alpha-0}) to build the ansatz:

\begin{equation}
\phi = 
\left\{
\begin{array}{lr}
A + B e^{-\Sigma}, & \mbox{if }\Sigma > -n \vspace{0.2cm} \\
0, &  \mbox{if }\Sigma \leq -n
\end{array}
\right.
\end{equation}
We define the following auxiliary functions:

\begin{equation}
G_k(n) = \int_{-n}^{\infty}e^{-k\Sigma}P_F(\Sigma) \ud\Sigma
\label{eq:auxiliary-functions}
\end{equation}
Note that $G_k$ is monotonically increasing for all $k$ and we have the following properties:
\[
\lim_{n\rightarrow \infty} G_0(n) = 1\quad,\quad\lim_{n\rightarrow \infty} G_1(n) = 1\quad,\quad\lim_{n\rightarrow \infty} G_2(n) = \infty\quad\mbox{and}\quad G_k(n)\leq e^{kn}\,\,\forall\,\,k\geq 0
\]
Imposing the constraints leads to the system
\begin{equation}
\left\{
\begin{array}{l}
AG_0(n) + BG_1(n) = \avg{\phi}_F
\vspace{0.2cm} \\
AG_1(n) + BG_2(n)=-\avg{\phi}_B \\
\end{array}
\right.
\label{eq:sys-alpha-0-alt}
\end{equation}
with solution
\begin{equation}
A = \frac{G_1(n)\avg{\phi}_B+G_2(n)\avg{\phi}_F}{G_0(n)G_2(n)-G_1(n)^2}\quad\quad\mbox{and}\quad\quad B = \frac{G_0(n)\avg{\phi}_B+G_1(n)\avg{\phi}_F}{G_1(n)^2-G_0(n)G_2(n)}
\label{eq:A-B}
\end{equation}
Finally the second moment is:
\begin{IEEEeqnarray}{rCl}
\nonumber
\avg{\phi^2}_{F,n} & = & A^2G_0(n) + 2ABG_1(n) + B^2G_2(n) \\[0.2cm]
& = & \frac{G_0(n)\avg{\phi}_B^2 + 2G_1(n)\avg{\phi}_B\avg{\phi}_F + G_2(n)\avg{\phi}_F^2}{G_0(n)G_2(n)-G_1(n)^2}
\label{eq:ap-second-moment}
\end{IEEEeqnarray}
and hence
\begin{equation}
\lim_{n\rightarrow\infty} \avg{\phi^2}_{F,n} = \avg{\phi}_F^2 \Rightarrow \lim_{n\rightarrow \infty} \var{\phi}_{F,n} = 0
\end{equation}
saturating the bound.

\section{Continuity and concavity of the bound with respect to $\alpha$}

For a given distribution $P_F(\Sigma)\neq \delta(\Sigma)$, the bound in eq (\ref{eq:TUR-alpha}) is given by the function $\mathcal{B}(\alpha)$:

\begin{equation}
\mathcal{B}(\alpha) \equiv \frac{\alpha}{2\avg{(\coth(\Sigma/2)+1 - 2\alpha)^{-1}}_F} -\alpha(1-\alpha)
\end{equation}
In this appendix, we show that this function is concave and continuous. As shown in appendix \ref{ap:deduction} and argued in the main text, for a fixed $\alpha$ one can always find a distribution $P_F^{(\alpha)}(\Sigma, \phi)$ that has $P_F(\Sigma)$ as a marginal and saturates the bound (or in problematic cases for $\alpha = 0, 1$ we can find a sequence $P_{F,n}^{(\alpha)}(\Sigma, \phi)$ of such distributions that tend to saturation). That is, defining $P_B^{(\alpha)}(\Sigma, \phi) = P_F^{(\alpha)}(-\Sigma, -\phi)e^{\Sigma}$ we have

\begin{equation}
(1-\alpha) \var{\phi}^{(\alpha)}_F + \alpha \var{\phi}^{(\alpha)}_B = \left(\avg{\phi}^{(\alpha)}_F + \avg{\phi}^{(\alpha)}_B\right)^2\mathcal{B}(\alpha)
\label{eq:bound-full}
\end{equation}
Since the bound $\mathcal{B}(\alpha)$ depends only on $P_F(\Sigma)$ and not on the constraints over $\phi$, we will choose the constraints to be such that $\avg{\phi}_F + \avg{\phi}_B = 1$, meaning eq (\ref{eq:bound-full}) becomes simply
\begin{equation}
(1-\alpha) \var{\phi}^{(\alpha)}_F + \alpha \var{\phi}^{(\alpha)}_B = \mathcal{B}(\alpha)
\label{eq:bound-simple}
\end{equation}
So let $\alpha, \beta, t \in [0,1]$ and $\gamma = t\alpha + (1-t)\beta$. If we consider the distribution $P_F^{(\gamma)}$ (or the limit of the $P_{F,n}^{(\gamma)}$) this leads to

\begin{IEEEeqnarray}{rCl}
\nonumber
\mathcal{B}(\gamma) & = & (1-\gamma) \var{\phi}^{(\gamma)}_F + \gamma \var{\phi}^{(\gamma)}_B \\[0.2cm] \nonumber
& = & (1-(t\alpha + (1-t)\beta)) \var{\phi}^{(\gamma)}_F + (t\alpha + (1-t)\beta) \var{\phi}^{(\gamma)}_B \\[0.2cm] \nonumber
& = & t\left((1-\alpha) \var{\phi}^{(\gamma)}_F + \alpha \var{\phi}^{(\gamma)}_B\right) + (1-t)\left((1-\beta) \var{\phi}^{(\gamma)}_F + \beta \var{\phi}^{(\gamma)}_B\right) \geq \\[0.2cm]
& \geq & t\mathcal{B}(\alpha) + (1-t)\mathcal{B}(\beta)
\end{IEEEeqnarray}
implying concavity. From the concavity of $\mathcal{B}$ it follows that it must be continuous in the interior of the interval $]0,1[$.

To show continuity in the extremities we will focus in $\alpha=0$ (that proves continuity for $\alpha=1$ by swapping the forwards and backwards processes). We consider once more the distributions $P_{F,n}$ defined in the appendix [\ref{ap:saturation}]. We already know from the previous appendices that the $P_{F,n}$ can asymptotically saturate the bound when $\avg{e^{-2\Sigma}}_F = \infty$. To see that this happens even when $\avg{e^{-2\Sigma}}_F<\infty$ we first note that eq (\ref{eq:ap-second-moment}) remains valid even in this case, so

\begin{equation}
\lim_{n\rightarrow\infty} \avg{\phi^2}_{F,n} = \lim_{n\rightarrow\infty} \frac{G_0(n)\avg{\phi}_B^2 + 2G_1(n)\avg{\phi}_B\avg{\phi}_F + G_2(n)\avg{\phi}_F^2}{G_0(n)G_2(n)-G_1(n)^2} = \frac{\avg{\phi}_B^2 + 2\avg{\phi}_F\avg{\phi}_B+\avg{\phi}_F^2\avg{e^{-2\Sigma}}_F}{\avg{e^{-2\Sigma}}_F - 1} = \avg{\phi}_F^2 + \frac{(\avg{\phi}_F + \avg{\phi}_B)^2}{\avg{e^{-2\Sigma}}_F - 1}
\end{equation}
which matches eq (\ref{eq:phi2-alpha-0}) implying the $P_{F,n}$ still get arbitrarily close to saturating the bound when $n\rightarrow\infty$. On the other hand, if we consider the corresponding backwards process $P_{B,n}(\Sigma,\phi) = P_{F,n}(-\Sigma, -\phi)e^{\Sigma}$, then

\begin{equation}
\var{\phi}_{B,n} = \avg{\phi^2}_{B,n} - \avg{\phi}_B^2 = \avg{\phi^2e^{-\Sigma}}_{F,n} - \avg{\phi}_B^2 = A^2G_1(n) + 2ABG_2(n) + B^2G_3(n) - \avg{\phi}_B^2
\end{equation}
where $A$ and $B$ are given by eq (\ref{eq:A-B}). Since $P_F(\Sigma)\neq \delta(\Sigma)$, then there exists $n_0$ such that $n>n_0$ implies $G_0(n)G_2(n) - G_1(n)^2 > 0$. On the other hand, the $G_k(n)$ must be finite, for finite $n$ and $k\geq 0$. The consequence is that as $n$ increases, $\var{\phi}_{F,n}$ tends to the bound while $\var{\phi}_{B,n}$ remains finite (but it may be infinite in the limit).

So suppose by absurd that $\mathcal{B}(\alpha)$ is discontinuous at $\alpha = 0$. Since $\mathcal{B}$ is concave this means that
\begin{equation}
L\equiv \lim_{\alpha\rightarrow 0} \mathcal{B}(\alpha) = \mathcal{B}(0) + \varepsilon
\label{eq:limit-violation}
\end{equation}
with $\varepsilon>0$. Since the $P_{F,n}$ tend to saturation, then there exists $N$ such that 

\begin{equation}
\var{\phi}_{F,N} = \mathcal{B}(0) + \varepsilon' < \mathcal{B}(0) + \varepsilon = L
\end{equation}
where $0<\varepsilon'<\varepsilon$ and $\var{\phi}_{B,N} \equiv V$ is finite. So

\begin{equation}
(1-\alpha) \var{\phi}_{F,N} + \alpha \var{\phi}_{B,N} = (1-\alpha)(\mathcal{B}(0) + \varepsilon') + \alpha V
\label{eq:violation-1}
\end{equation}
for every $\alpha\in [0,1]$. So if we consider an $\alpha$ that is positive but arbitrarily small, we'd have $\mathcal{B}(\alpha) = L+\varepsilon''$, where $|\varepsilon''|$ can be made arbitrarily small by doing $\alpha\rightarrow 0$. Since

\begin{equation}
(1-\alpha) \var{\phi}_{F,N} + \alpha \var{\phi}_{B,N} \geq \mathcal{B}(\alpha)
\label{eq:violation-2}
\end{equation}
this implies, combining eqs (\ref{eq:limit-violation}) (\ref{eq:violation-1}) and (\ref{eq:violation-2})

\begin{equation}
(1-\alpha)(L - \varepsilon + \varepsilon') + \alpha V \geq L+\varepsilon'' \Leftrightarrow \alpha(V - L + \varepsilon - \varepsilon') \geq \varepsilon'' + \varepsilon - \varepsilon'
\label{eq:violation-3}
\end{equation}
Taking $\alpha$ sufficiently small, we get $|\varepsilon''| < \nicefrac{(\varepsilon - \varepsilon')}{2}$ and hence

\begin{equation}
\Leftrightarrow \alpha(V - L + \varepsilon - \varepsilon') > \frac{\varepsilon - \varepsilon'}{2}
\label{eq:violation-4}
\end{equation}
But the right hand side of this inequality is positive and independent of $\alpha$, while the left hand side is $\alpha$ times a value that is independent of $\alpha$. So for a sufficiently small $\alpha$ the inequality in (\ref{eq:violation-4}) is violated, leading to a contradiction.

\section{Saturation for distributions with two points in the support}
\label{ap:two-point}

Suppose that we have a distribution $P(\Sigma, \phi)$, obeying the fluctuation theorem $\avg{e^{-\Sigma}} = 1$ and with only two points in its support, namely $(\Sigma_1, \phi_1)$
and $(\Sigma_2, \phi_2)$.

From what we already derived, we can show $P$ saturates the bound by showing that the support obeys equation (\ref{eq:opt-support}) in the main text. In this case, this reduces to finding $\lambda$ and $\mu$ such that

\begin{equation}
\left\{
\begin{array}{l}
\lambda + \mu e^{-\Sigma_1} = \phi_1(1 -\alpha + \alpha e^{-\Sigma_1})\\
\lambda + \mu e^{-\Sigma_2} = \phi_2(1 -\alpha + \alpha e^{-\Sigma_2})
\end{array}
\right.
\end{equation}
which is a linear system with $\lambda$ and $\mu$ as its variables. Since the determinant of this system is $e^{-\Sigma_2} - e^{-\Sigma_1}$, there is always a solution when $\Sigma_1 \neq \Sigma_2$, meaning the bound is saturated.

If on the other hand $\Sigma_1 = \Sigma_2$, then $\avg{e^{-\Sigma}} = 1$ implies $\Sigma_1 = \Sigma_2 = 0$. In this case we have $P_F(\Sigma) = \delta(\Sigma)$ and the bound becomes undefined (see appendix \ref{ap:saturation}).
\end{document}